\documentclass{emulateapj} 
\newcommand{\uv}{\mbox{$u$-$v$}}
\newcommand{\ex}[1]{\mbox{$\times 10^{#1}$}}

\newcommand{\kms}{\mbox{km s$^{-1}$}}

\newcommand{\Jb}{\mbox{Jy bm$^{-1}$}}
\newcommand{\pyear}{\mbox{\% yr$^{-1}$}}
\newcommand{\Wmsr}[1]{\mbox{$\, \times ~10^{#1}$} {\mbox W~m$^{-2}$ Hz$^{-1}$ sr$^{-1}$}}

\newcommand{\SigG}{\mbox{$\Sigma_{\rm 1 GHz}$}}

\newcommand{\Ra}[4]{\mbox{${#1}^{\rm h} \; {#2}^{\rm m} \; {#3}\fs{#4} $}}
\newcommand{\dec}[4]{\mbox{${#1}\arcdeg \; {#2}\arcmin \; {#3}\farcs{#4} $}}

\shortauthors{Bietenholz}
\shorttitle{Radio Images of 3C~58}

\begin{document}
      
\title{Radio Images of 3C~58: Expansion and Motion of its Wisp}

\author{M. F. Bietenholz}
\affil{Department of Physics and Astronomy, York University, Toronto, M3J~1P3, Ontario, Canada}

\author{\em Accepted for publication in the Astrophysical Journal} 
 
\begin{abstract}
New 1.4~GHz VLA observations of the pulsar-powered supernova remnant
3C~58 have resulted in the highest-quality radio images of this object
to date.  The images show filamentary structure over the body of the
nebula.  The present observations were combined with earlier ones from
1984 and 1991 to investigate the variability of the radio emission on
a variety of time-scales.  No significant changes are seen over a
110~day interval.  In particular, the upper limit on the apparent
projected velocity of the wisp is 0.05~$c$.  The expansion rate of the
radio nebula was determined between 1984 and 2004, and is $0.014 \pm
0.003$\pyear, corresponding to a velocity of $630 \pm 70$~\kms\ along
the major axis.  If 3C~58 is the remnant of SN~1181, it must have been
strongly decelerated, which is unlikely given the absence of emission
from the supernova shell.  Alternatively, the low expansion speed and
a number of other arguments suggest that 3C~58 may be several thousand
years old and not be the remnant of SN~1181.
\end{abstract}


\section{INTRODUCTION}
\label{sintro}

The supernova remnant \objectname[]{3C~58} (\objectname[]{G130.7+3.1})
was classified as a pulsar wind nebula \citep[PWN, also
plerion; e.g.,][] {WeilerP1978}
long before a pulsar was known.  Its pulsar, \objectname[]{PSR
J0205+6449}, was only recently detected, first in the X-ray
\citep{Murray+2002} and then in the radio \citep{Camilo+2002}.
3C~58 is at a distance, $D$, of $\sim$3.2~kpc
\citep{Roberts+1993}\footnote{Note that a distance slightly larger
than this is possible, and that the pulsar dispersion measure is
approximately twice what is expected at that distance
\citep{Camilo+2002}.}.
3C~58 is a bright radio source, and previous radio imaging
observations of the synchrotron nebula have shown it to have a
center-brightened morphology, and a size of $6' \times 9'$
\citep{WeilerS1971, WilsonW1976, Green1986, ReynoldsA1988, 3C58-2001}.

The pulsar in the center of the nebula is loosing rotational energy
due to magnetic braking, at a rate which can be calculated from the
pulsar's spindown rate, and is $2.7 \times 10^{37}$~erg~s$^{-1}$
\citep[assuming a neutron star moment of inertia of
$10^{45}$~g~cm$^2$;][]{Murray+2002, Ransom+2004}.  This is the third
highest spindown power of any Galactic pulsar, the highest being that of
\objectname[]{PSR B0531+21} in the \objectname[]{Crab Nebula}.
This energy flows from the pulsar as a wind consisting of relativistic
particles and magnetic field, which then powers the visible
synchrotron nebula.  The details of this energy flow, however, are not
yet well understood.  In the Crab Nebula, which is the best studied
pulsar wind nebula, rapid variability has been seen throughout the
region near the pulsar at radio, optical, and X-ray wavelengths.  In
particular, a series of features called wisps propagate outward at
projected speeds of order $0.3\,c$.  The Crab nebula wisps are visible
in the radio \citep{Crab-2001, Crab-2004}, in the optical
\citep[e.g.,][]{Hester+1996, TanvirTT1997}, and in the X-ray
\citep{Mori+2002}.

These wisps are thought to be associated with the termination shock in
the pulsar's outflow, and indeed constitute almost the only
observational window on that outflow before it enters the body of the
nebula.  Radio features apparently associated with the pulsar outflow
are visible in pulsar nebulae other than the Crab, for example in
\objectname[]{Vela} \citep{Vela-wisp, Bock+2002}. In particular,
\citet{FrailM1993} showed that 3C~58 also has a radio feature with a
wisp-like morphology.  It would clearly be of interest to determine
whether 3C~58's wisp moves as rapidly as the wisps in the Crab, and to
determine whether rapidly moving wisps are a general feature of young
pulsar wind nebula, or a peculiarity of the Crab.

Another outstanding question concerning 3C~58 is its expansion rate,
in particular that of the synchrotron bubble.  3C~58 has often been
associated with the supernova of 1181
A.D. \citep[SN~1181;][]{ClarkS1977, StephensonG1999}, implying an age
of $\sim 820$~yr for the nebula.  Such a young age suggests that the
nebula should be expanding rapidly.  The measured expansion speeds,
however, of both the synchrotron bubble \citep{3C58-2001} and of the
thermal filaments \citep{Fesen1983, FesenKB1988, vdBergh1990}, seem to
be considerably lower than expected for an age of $\sim 820$~yr.

Such slow current expansion speeds might be observed if 3C~58's
expansion has been greatly decelerated, but this is unlikely for the
following reason.  The synchrotron nebula is expected to be expanding
into the supernova ejecta, which are themselves expanding.  The
synchrotron nebula is at fairly high pressure compared to that
expected in the supernova ejecta, so in order for synchrotron bubble
to have decelerated, the ejecta would also have to be decelerated.  If
the ejecta had been decelerated, however, one would expect
bright radio emission from their interaction with the surrounding
material.  Such emission has not been observed despite numerous
searches. \citep[For more details on this argument, see][]{3C58-2001}.
In addition, such deceleration of the ejecta would likely produce
considerably more thermal X-ray emission than is in fact seen
\citep{Slane+2004}.  The low measured expansion velocities suggest
that 3C~58 might be several thousand years old, and therefore not be
the remnant of SN~1181.  An accurate determination of the expansion
speed of the synchrotron bubble is therefore important to
reconstructing the history of 3C~58, and determining whether it is in
fact associated with SN~1181.

I obtained new radio observations of 3C~58 using the
NRAO\footnote{The National Radio Astronomy Observatory, NRAO, is
operated under license by Associated Universities, Inc., under
cooperative agreement with National Science Foundation.} Very Large
Array (VLA), with the goals of addressing both the question of
possible rapid motions near the pulsar, and in order to obtain an
accurate measurement of the present expansion speed of the synchrotron
nebula. I describe the observations in \S~\ref{sobs},
discuss the results in \S~\ref{sresults}, and discuss
my findings in \S~\ref{sdiscuss}.

\section{OBSERVATIONS AND DATA REDUCTION}
\label{sobs}

I observed 3C~58 using the A, B, and C configurations of the VLA at
1.4~GHz.  Details of the observing runs are given in
Table~\ref{tobs}. Two separate A-configuration runs, separated by
110~d, were obtained in order to check for rapid changes at the
highest resolution.  To avoid bandwidth smearing, the observations
were carried out in spectral line mode, with a total bandwidth of
44~MHz comprising 7 spectral channels, each of width 6.25~MHz.  The
data were calibrated and imaged using NRAO's AIPS software package.
The flux density scale and instrumental bandpass were determined using
observations of 3C~286.

I also re-edited the data of \citet{ReynoldsA1988}, taken in 1984 and
also using the A, B, and C array configurations at 1.4~GHz, but
without using the spectral line mode and having a bandwidth of
12.5~MHz.  These data were then imaged and de-convolved in a similar
manner as those from 2003/2004.  Finally, I also re-reduced archival
A-configuration 5~GHz VLA observations of 3C~58 taken on 1991 June 30,
with a bandwidth of 50~MHz \citep[original results published
by][]{FrailM1993}.

Obtaining high-dynamic-range images of 3C~58 presents a challenge
because the field around 3C~58 contains a number of background
sources, including a nearby bright extra-galactic double.  To obtain
the best image of 3C~58, these background sources must be included in
the deconvolution, otherwise their sidelobes will degrade the image of
3C~58.  Although maximum entropy deconvolution is well suited for
extended emission like 3C~58, the program available in AIPS (task
VTESS) is not capable of treating the multiple fields necessary to
also deconvolve the background sources.  Furthermore, the
deconvolution of compact objects like the background sources is better
carried out using CLEAN rather than maximum entropy deconvolution.  I
therefore used a hybrid deconvolution scheme which made use of both
maximum entropy and CLEAN, with the final images made as follows.
First, a CLEAN deconvolution was performed on both the 3C~58 field and
of six secondary fields to include the nearby background sources.
Such multi-field CLEAN images were used to fully self-calibrate the
data in both amplitude and phase.  For the final images of 3C~58, I
subtracted the CLEAN components for all the background sources, in
other words all the clean components except those for 3C~58, from the
fully self-calibrated \uv~data.  The resulting \uv~data set was then
imaged and deconvolved using VTESS\@.  The VTESS (maximum entropy)
deconvolution reduces the rippling at small spatial scales which was
visible in the CLEAN images of 3C~58 and is a known instability of the
CLEAN algorithm.

In order to determine whether there is any rapid variability in
3C~58's radio emission, I made separate images from my 2003 July 7 and
2003 August 9 data.  Time resolved radio observations of 3C~58 with
the VLA, however, present the problem that it exhibits structure at
spatial scales from $\sim$10\arcmin\ down to $<$1\arcsec.  Such a wide
range of spatial scales is not well sampled by any single VLA array
configuration.  The A array configuration at 1.4~GHz, for example,
samples only spatial scales between $\sim 1.4\arcmin$
and $\sim$1\farcs2.  The usual approach of recovering the
large scale structure by additional observations using more compact
array configurations is precluded by the fact that VLA configuration
changes occur only every four months.

I used a strategy for obtaining reliable, time-resolved images of an
extended object similar to that developed by \citet{Crab-2001}.  Since
a speed of $c$\/ represents a proper motion of only $\sim$1\farcs6 per
month,
any rapid evolution must occur on the smaller spatial scales, which
are well sampled by the A~array configuration.  The large scale
structure as sampled by the B and C array configurations (spatial
scales of $5\arcsec \sim 15\arcmin$), therefore, is unlikely to
change on timescales of $\sim 2$~months.
I therefore made separate images of each A~array session by supplying
the maximum entropy deconvolution with a default image. The default
image was made by imaging all the 2003/2004 data, including that from
both A, and the C and D array configurations, and subsequently
smoothing the image by convolving with a Gaussian of FWHM 30\arcsec.
The use of the same default for both A-configuration images will also
serve to minimize any spurious differences between them since any
differences of each from the common default will be only such as are
demanded by the data.

\section{RESULTS}
\label{sresults}

\subsection{Images}
\label{simages}

\begin{figure}
\centering
\includegraphics[width=1.10\textwidth]{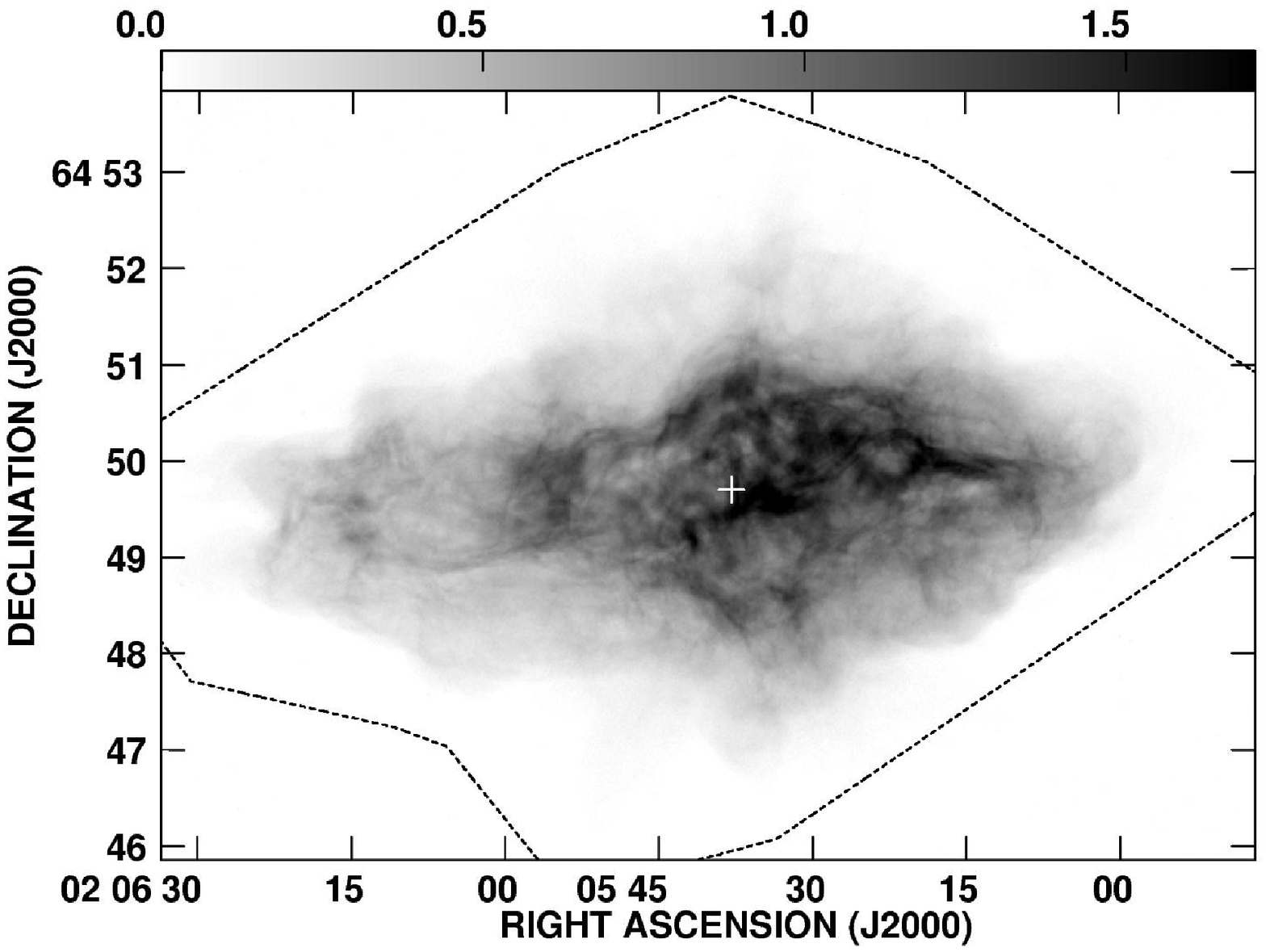} 
\caption{An image of 3C~58 at 1.4~GHz, made using all the A, B and C
array-configuration VLA data from 2003 and 2004.  The FWHM of the Gaussian
convolving beam was 1\farcs36.  The cross marks the position of the
pulsar, PSR~J0205+6449 \citep[RA = \Ra{02}{05}{37.92}, decl.\ =
\dec{64}{49}{42}{8};][]{SlaneHM2002, Camilo+2002}, which is known to
about 0\farcs5.  The grey-scale is labelled in m\Jb, and the rms of
the background was $\sim 12 \; \mu$\Jb. The background sources,
including the bright extragalactic double at R.A. = \Ra{2}{6}{31}{7},
decl.\ = \dec{64}{54}{14}{5}, were subtracted from the visibility data
before the final maximum-entropy deconvolution, hence are not visible
in this image (see text \S~\ref{simages} for details). The dashed line
indicates the fitting region used for the expansion calculation (see
\S~\ref{sexp}).  For a detail of the region near the pulsar, see
Figure~\protect\ref{fwisp}.}
\label{ffull}
\end{figure}

\begin{figure}
\centering
\includegraphics[width=\textwidth]{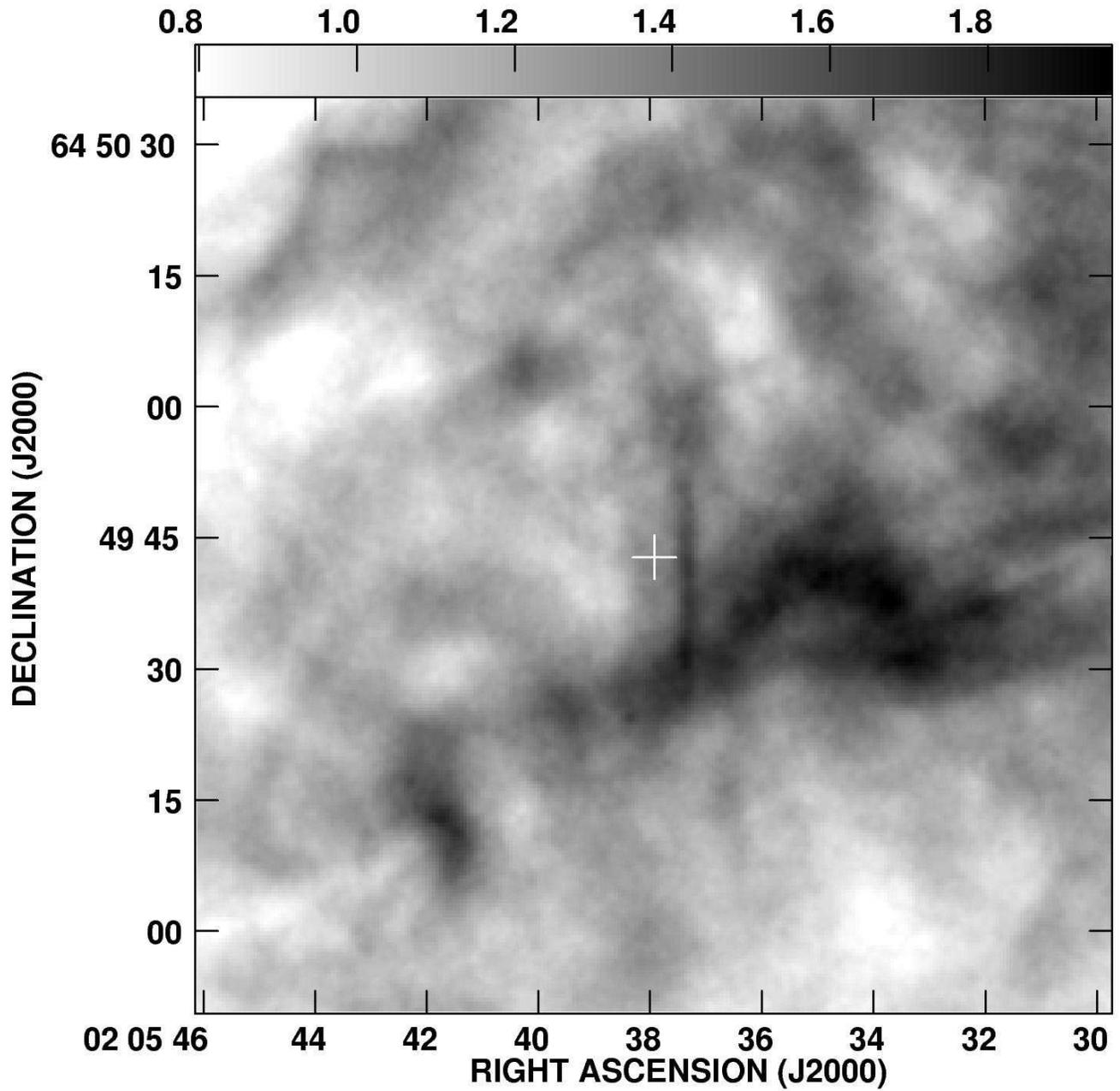} 
\caption{An image of the region near the pulsar in 3C~58.  This figure
shows the central part of the image in Figure~\protect\ref{ffull}. The
cross again marks the position of the pulsar, known to $\sim
0.5\arcsec$.}
\label{fwisp}
\end{figure}

The full resolution image of 3C~58 is shown in Figure~\ref{ffull}.
This image was made by combining all the data in Table~\ref{tobs},
including both A array configuration sessions.  It therefore
represents a time-average of the emission of the observing period of
2003 July to 2004 April.  The position of the recently discovered
pulsar, taken from \citet{SlaneHM2002} and \citet{Camilo+2002}, is
marked.  The image was made using robust weighting, with the AIPS
robustness parameter set to 0, and the Gaussian convolving beam had a
FWHM of 1\farcs36.
In addition, Figure~\ref{fwisp}, shows a detail of the central
region, including the ``wisp'' identified by \citet{FrailM1993}.  
I now turn to examining the variability of 3C~58's radio emission on
several timescales.

\subsection{Expansion between 1984 and 2003}
\label{sexp}

I examine first the longest timescales in order to determine the
overall expansion rate.  This can be most accurately determined from
the pair of multi-configuration radio images from 1984 and
2003/2004\footnote{The image made from the 1991 data is not suitable
for this purpose, firstly because it consists only of A-configuration
data, hence does not reliably image structure larger than 1.4\arcmin,
and secondly because it suffers from bandwidth smearing at distances
$>45\arcsec$ from the phase-center \citep[see e.g.,][]{Cotton1999,
BridleS1999}. This smearing can cause small but systematic radial
shifts of the position of brightness peaks due to the non-symmetrical
VLA bandpass.}.  I use the same approach to determining the expansion
as was used in \citet{3C58-2001}, and repeat a brief description here
for the convenience of the reader.  The goal is to determine the
overall or average expansion speed of the radio nebula.  Since there
are few well-defined, compact features, the expansion is measured not
by determining the proper motion of individual features, but by
determining an overall scaling between a pair of images by
least-squares. This was accomplished by using the MIRIAD task
IMDIFF\footnote{The IMDIFF program was slightly modified so as to
treat blanking correctly.} which determines how to make one image most
closely resemble another by calculating unbiased estimators for the
scaling in size, $e$, the scaling and the offset in flux density, $A$
and $b$ respectively, and the offsets in RA and decl., $x$ and $y$
respectively, needed to make the second image most closely resemble
the first.  My chief interest is in the expansion factor, $e$, but
because of uncertainties in flux calibration, absolute position, and
image zero-point offsets caused by missing short spacings, all five
parameters were determined.  This method was originally developed by
\citet{TanG1985} and more details of its use in a similar situation
are given by \citet{Crab-expand}.  Before running IMDIFF, the
2003/2004 image was convolved to the resolution of the 1984 one,
namely $2\farcs00 \times 2\farcs07$ (FWHM) at p.a.\ $-77$\arcdeg.  The
region exterior to 3C~38 was blanked so as to exclude it from the
fitting.  The fitting region encompassed 37.3 square arcminutes or
$\sim$46,300 beam areas, and is indicated by a dotted line in
Figure~\ref{ffull}.  The resulting expansion factor between the 1984
and the 2003/2004 images is 1.0027, with a formal uncertainty of $\sim
0.0001$.

Although the formal uncertainty in $e$ takes into account correlations
of $e$ with the other parameters ($A, b, x$ and $y$), it is still
likely a significant underestimate because the true uncertainty will
be dominated by systematic effects rather than by random noise.
Systematic effects likely to influence my determination of $e$ are
problems in recovering the largest scale structure in the
deconvolution, bandwidth smearing, and the primary beam correction. I
discuss each in turn.

There are often problems in recovering the largest scale structure
from interferometric observations such as these, and such problems
might bias my determination of $e$.  To eliminate this possibility, I
high-pass filtered my images using a Gaussian of FHWM 30\arcsec, and
obtained a marginally smaller expansion factor of 1.0022.  To check
whether biases are introduced by possible edge-effects, I determined
$e$ using only pixels above a certain cutoff in brightness.  I
determined $e$ using the pixels with brightness $>3$\%\ and $>20$\%\
of the peak, resulting in fitting areas of 90\% and 52\% of the above
value, and values of $e$ of 1.0021 and 1.0025 respectively.
Similarly, tests show that the value of $e$ is not sensitive to
whether CLEAN or VTESS deconvolution is used. I conclude that my
determination of $e$ is not sensitive to deconvolution errors at a
level greater than 0.0006.

I note that the total flux density of 3C~58 was found to be increasing
with time at a rate of about 0.28\pyear\ \citep{Green1987,
AllerR1985}., which would imply an increase of $\sim 5.5$\% in the
total flux density between 1984 and 2003/2004. I find that the 5~GHz
flux density of 3C~58 {\em decreased}\/ by $0.7 \pm 8.0$\% between
1984 and 2003/2004 which is consistent with the earlier measured
secular increase but does not confirm it. 

The data of 1984 had a bandwidth of 12.5~MHz, and as a result will be
subject to a small amount of bandwidth smearing at distances $\gtrsim
200$\arcsec\ from the image center.  Bandwidth smearing will smear the
image radially.  Such radial smearing might possibly bias the derived
value of $e$, although to first order it preserves surface brightness
\citep[see, e.g.,][]{Cotton1999}.  As a test, I averaged the 2003/2004
data over 25~MHz in frequency (from the original channel width of
6.25~MHz).  Using IMDIFF to compare this artificially
bandwidth-smeared image to the unsmeared original, I find the value of
$e$ to differ from unity by $<0.0003$.
As the amount of bandwidth smearing in this test was twice that in the
1984 data set, I conclude that any effect of bandwidth smearing of
the 1984 data set is small (and would cause my estimate of the
expansion rate to be slightly but not significantly {\em
over}-estimated).

Finally, different primary beam corrections might cause an error in
the determination of $e$.  I obtained my IMDIFF results from images
uncorrected for the primary beam pattern of the VLA, which has the
desirable result of making the noise in the image spatially uniform.
If the two images were taken at a different observing frequency, the
primary beam correction would differ, and a bias might be introduced.
Since, however, the primary beam correction is not large, (being only
$\sim 7$\% near the eastern and western edges of the nebula), and
since the 1984 and 2003/2004 images are at almost the same frequency
(the center frequencies differ by only 21~MHz), the difference in the
primary beam correction between the 1984 and 2003/2004 images is
negligible.  As a test, IMDIFF was run between images made separately
from the two intermediate frequencies in the 2003/2004 data set, which
differ in frequency by a larger amount of 80~MHz.  The resulting 
change in $e$ was only 0.0005.

As mentioned above, the uncertainty in $e$ is likely dominated by
difficult-to-determine systematic effects.  The tests performed
above, however, in no case resulted in a change $> 0.0006$ in the
derived value of $e$, and I will adopt this value as a conservative
uncertainty in $e$.

The value of $e$ implies 3C~58's rate of expansion over the period
1984 to 2004 is $0.014 \pm 0.003$\pyear.
This value is consistent with, but more accurate than, the one of
$0.020 \pm 0.008$\pyear\ obtained by \citet{3C58-2001} from a variety
of earlier, lower-resolution radio images.
I note here that this expansion rate is much smaller than the rate of
0.124\pyear\ which would be expected for undecelerated expansion of
3C~58 since 1181~A.D.

\subsection{Changes between 1991 and 2003/2004}
\label{s91diff}

Are there any changes in 3C~58's morphology apart from the relatively
slow expansion?  The Crab Nebula shows complex brightness changes,
distinct from the general expansion, which occur over a timescale of a
few years \citep{Crab-2004, Crab-2001, Crabwisp-1992}.  Is there
similar variability in 3C~58?  The best pair images for addressing
this question are those from 1991 and 2004, since they have the
highest resolution and good signal-to-noise.  In 1991, only A
configuration observations are available, so only structure $\lesssim
1.4\arcmin$ in size can be imaged.  This is not a serious limitation
because any changes will likely relatively localized.
A more serious limitation is that the bandwidth of 43~MHz used in 1991
results in a considerable amount of bandwidth smearing --- features
far from the phase center will suffer radial smearing
45\arcsec\ of the phase center (see \S~\ref{sexp} above), which will
limit the search for motions or changes to the region near the pulsar.
The resolution of 1\farcs36 and the time interval of 12.4~yr means
I am sensitive to motions with projected speeds of $\gtrsim
1500$~\kms.

I formed a difference image by subtracting the 2003/2004 image from
that of 1991.  To account for any possible small shifts in nominal
position or differences in the amplitude scale etc., IMDIFF was first
run to determine the best-fit flux density scaling and offset and RA
and decl.\ offsets ($A, b, x,$ and $y$) between 1991 and 2003/2004
images using only the portions within $\pm 45\arcsec$ of the pulsar.
The scaling in size ($e$) cannot be reliably determined from such a
small region, so it was fixed at $e = 1.0017$, arrived at by assuming
the expansion rate of $0.014 \pm 0.003$\pyear\ derived in
\S~\ref{sexp} above.  After applying the relevant shifts and scalings,
I formed the difference image.  Because of both the IMDIFF scaling and
the use of the same default image, this difference image should
represent the smallest set of differences allowed by the data.

\begin{figure}
\centering
\includegraphics[width=\textwidth]{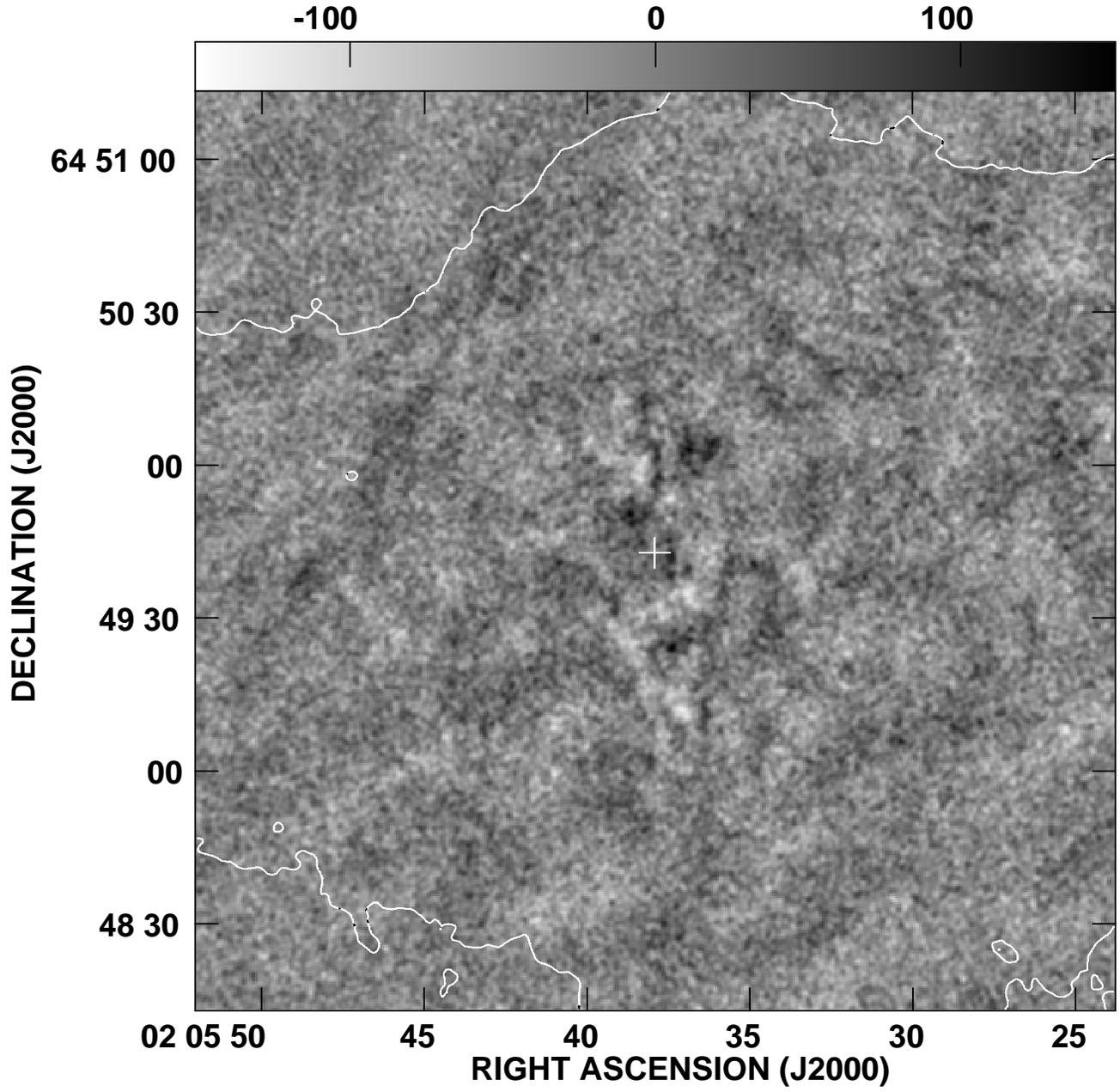} 
\caption[]{Difference between the images from 2003/2004 and 1991,
after the best-fit expansion, translation and scaling in brightness
(see text \S~\ref{s91diff}).  Due to the bandwidth smearing in 1991
image, the difference image becomes less reliable beyond $\sim
45$\arcsec\ from the pulsar position, which is marked with a cross.
The FWHM of the Gaussian convolving beam was 1\farcs36.  The contour,
included for reference, is the 30\% contour of the 2003/2004 image in
Figure~\ref{ffull}.  The grayscale is labelled in $\mu$\Jb.}
\label{fdiff91}
\end{figure}

This difference image is shown in Figure~\ref{fdiff91}. There is a
tantalizing hint of structure in the region near the pulsar.  The
extrema within 20\arcsec\ of the pulsar are +150 and $-120 \; \mu$\Jb,
or +8\% and $-6$\% of the peak brightness.  These extrema represent
$\sim5\sigma$, slightly larger than the extremes expected from a
purely random variation over the whole image (which comprises $n \sim
170,000$ beam areas).  However, the distribution of image-plane noise
is probably not strictly Gaussian, and the statistical significance of
the extreme values in the difference image is therefore likely
somewhat different than calculated from Gaussian statistics.  For this
reason, the evidence for brightness variations near the center of
3C~58 should be regarded as somewhat tentative.

\subsection{Variability over 110~Days}
\label{svar110}

\begin{figure}
\centering
\includegraphics[width=\textwidth]{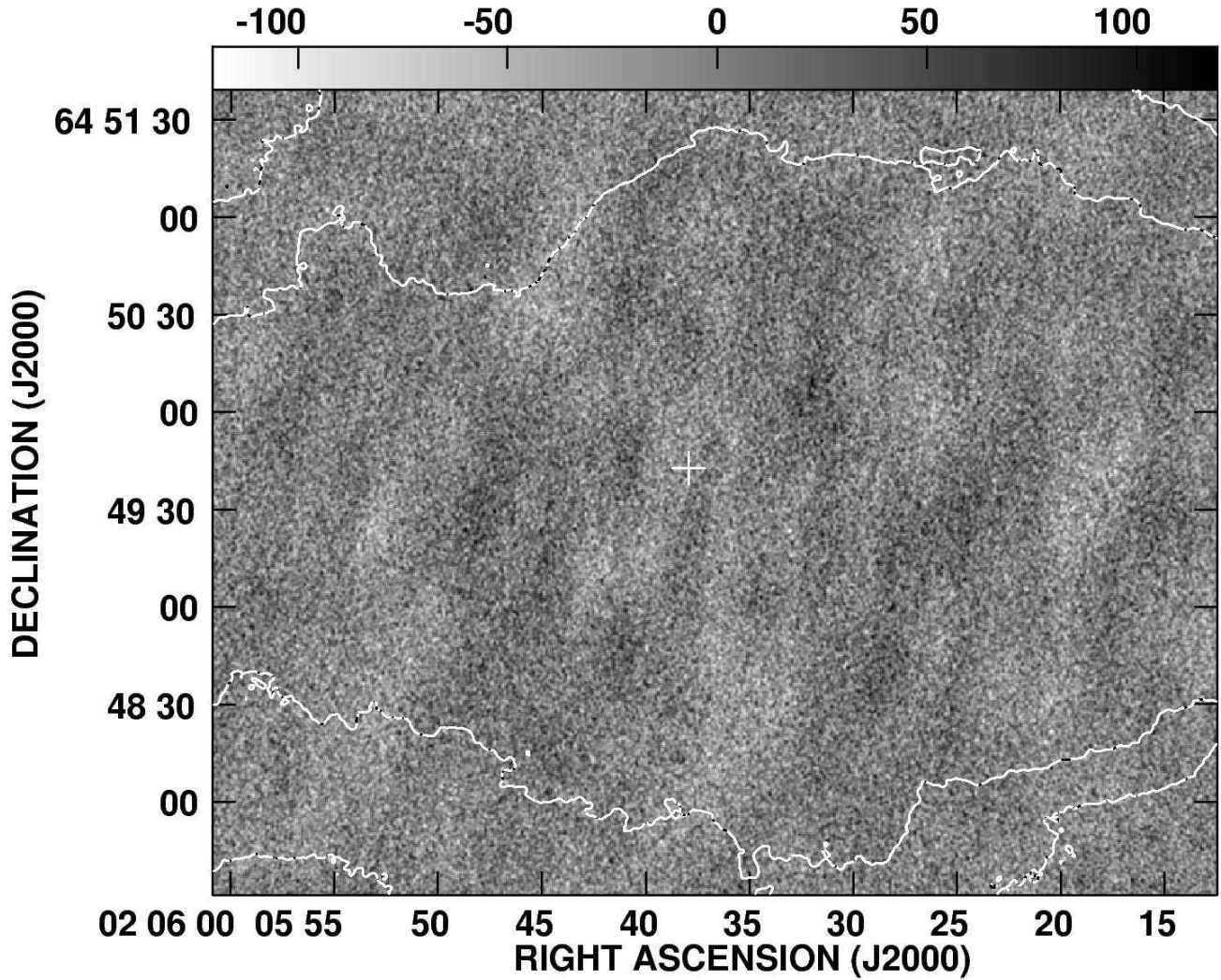}
\caption{Central portion of the difference image between 2003 July 7
and August 9, i.e. between images made using separately each of the
2003 A~configuration data sets.  The FWHM of the Gaussian convolving
beam was 1\farcs36.  The rms variation in the difference image is $25
\; \mu$\Jb.  The plotted contours are the 2 and 20\% contours from the
total intensity image in Figure~\protect\ref{ffull}, shown for
reference.  The position of the pulsar is marked by a cross.}
\label{fdifimg}
\end{figure}

Finally, I turn to determining whether there is any variability on the
short time-scales of less than 1~year.  For this purpose, I examine
the differences between the two images made using each of the 2003
A~configuration observations separately.  In Figure \ref{fdifimg}, I
show a difference image formed by subtracting image from July 7 from
that of August~9.  No structure above the noise is visible in the
difference image.  Any changes in the radio morphology of 3C~58 over
this period, therefore, were no larger than the noise. The rms
brightness of the difference image over 3C~58 is 26~$\mu$\Jb, which is
consistent with that expected from the
background rms of the individual images.  The rms brightness over the
region within $\pm 20\arcsec$ of the pulsar is not significantly
larger.  A limit of $\sim 160 \; \mu$\Jb, or $\sim 8$\% of the peak
brightness can be set any changes over the period of 110~d\footnote{As
noted earlier, the computation of the exact statistical significance
based on Gaussian statistics is likely to be in error.  The given
limits on the brightness variation over 110~d represent $\sim 5$ times
the rms in the difference image.  Tests performed by adding random
noise to the difference images of the Crab from \citet{Crab-2004} show
that the real brightness variations in the Crab are clearly
distinguishable from the noise at this level.}.

\begin{figure}
\centering
\includegraphics[height=4.5in]{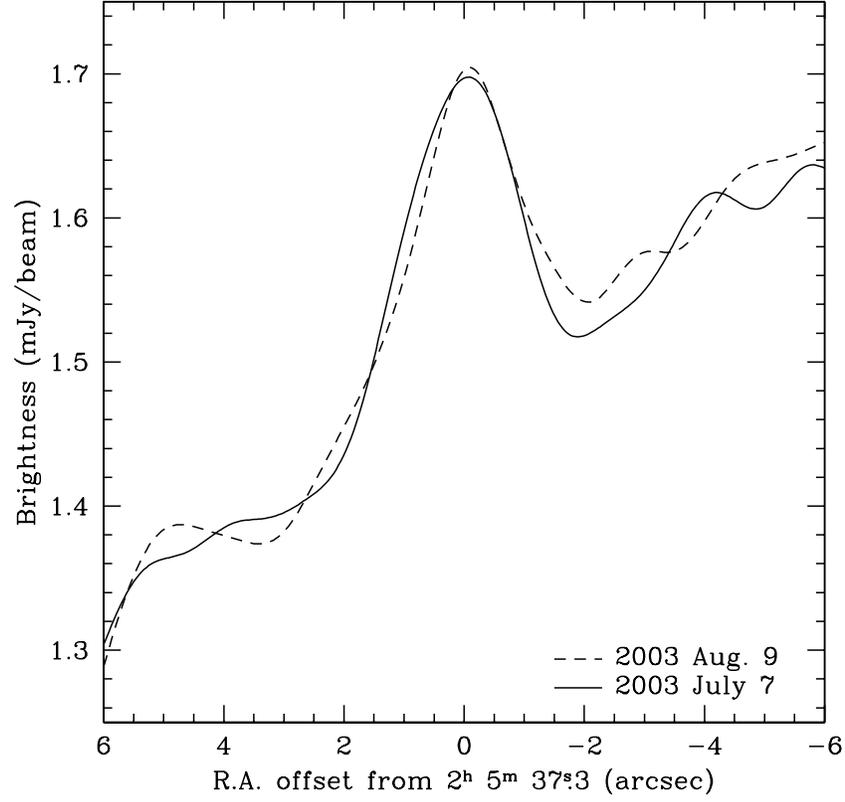} 
\caption{Profiles in R.A. through the ``wisp'' on 2003 July 7 and 9.
The profiles are at a declination of \dec{64}{49}{40}{2} (J2000).  To
improve the signal-to-noise ratio, the images were smoothed in the
declination direction by boxcar averaging over 2\farcs4.  The FWHM
Gaussian resolution in the R.A. direction is 1\farcs36. The
uncertainty is $\sim 0.015$~m\Jb\ in each profile.}

\label{fwispprof}
\end{figure}

In particular, the feature identified as a ``wisp'' by
\citet{FrailM1993} does not show any motion over this 110~d interval.
I show profiles drawn in the R.A. direction through the wisp in
Figure~\ref{fwispprof}.  To decrease the uncertainty, the profiles
were averaged over 2\farcs4 in declination.  Any changes between
the two epochs are smaller than the profile uncertainty of $\sim 15 \;
\mu$\Jb.
In particular, the wisp moves $< 0\farcs 25$ between these two epochs,
implying a projected speed of $\lesssim \, 0.05 c$ or $\lesssim
15,000$~\kms\ ($D = 3.2$~kpc).

\section{DISCUSSION}
\label{sdiscuss}

I produced new radio images of the supernova remnant 3C~58\@.  These
new images are the highest quality radio images of 3C~58 so far, By
comparing images made at intervals of $\sim$20~yr, $\sim$12~yr and
110~d, I was able to measure the expansion rate of the synchrotron
nebula, and to establish upper limits on any variations or motions
within the nebula.

The new high-resolution images show a general radio structure of 3C~58
very similar to that seen in earlier images at lower resolution and
signal-to-noise \citep[e.g.,][]{ReynoldsA1988}.  Filamentary structure
can be seen throughout the body of 3C~58, and the general texture is
very similar to both that seen in the X-ray images of 3C~58
\citep{Slane+2004} and in radio images of the Crab
Nebula \citep[e.g.,][]{Crab-2004}.  Several elongated loop-like
structures are visible, particularly to the East.  The narrowest
filaments are likely to be unresolved at my 1\farcs36 resolution, but
are relatively faint.  The brighter filaments seem to have widths of a
few arc-seconds.  The most prominent filament is one near the pulsar,
oriented North-South, which was identified earlier as a wisp
\citep{FrailM1993}.  I will discuss it separately in
\S~\ref{swispdiscuss} below.

\subsection{The Expansion Rate and Age of 3C~58}

By comparing my observations to ones taken in 1984, I measured the
current expansion rate of the radio nebula to be $0.014 \pm
0.003$\pyear, corresponding to speeds $630 \pm 70 \, (D/3.2$kpc)~\kms\
along the major axis of the nebula and approximately half that along
the minor axis\footnote{These speeds are calculated assuming
symmetrical expansion about the center of the nebula.  The
observations constrain the expansion center to lie within about
1.5\arcmin\ from the pulsar position.  The degree of one-sided
expansion is not well determined.}.
This measurement of the expansion rate is consistent with, but
considerably more accurate than the earlier one of \citet{3C58-2001}.
As already pointed out in that reference, the low expansion speed of
the radio nebula is corroborated by the slow proper motions of optical
filaments \citep{FesenKB1988,vdBergh1990},
and by the low radial velocities implied by the low optical
line-widths \citep[$\lesssim$900~\kms,][]{Fesen1983}.

All these measures of expansion speed of 3C~58 are much smaller than
that expected if it were the undecelerated remnant of SN~1181, in
which case the expansion rate would be 0.124\pyear\ or $\sim
5500$~\kms\ along the major axis.  Conversely, my measured expansion
rate suggests an age for 3C~58 of $\sim 7000$~yrs with a $3\sigma$
lower limit of 4300~yrs, which is incompatible with the usual
association with SN~1181.

If 3C~58 is in fact the remnant of SN~1181, then, it must have been
strongly decelerated.  As argued in \cite{3C58-2001}, to decelerate
the synchrotron nebula, whose pressure is likely higher than that in
the surrounding supernova ejecta, requires the ejecta to have been
even more strongly decelerated.  If the ejecta indeed had been
strongly decelerated, shell emission would be expected from their
interaction with the interstellar medium.  No shell emission, however,
is seen in 3C~58 in the radio despite several searches
\citep{3C58-2001, ReynoldsA1985}. 

The simplest way out of this dilemma is to suggest that 3C~58 is not
in fact the remnant of SN~1181, but rather is the remnant of a
considerably older supernova.  Several other lines of evidence suggest
that 3C~58 might be rather older than 820~yr \citep[many of these
points have already been discussed in][]{Chevalier2004,
Chevalier2005}.
\begin{trivlist}

\item{1.} The internal energy of the pulsar nebula, estimated by the
synchrotron minimum energy, is larger than the current spindown
luminosity multiplied by 820~yr \citep{Chevalier2004, Chevalier2005}.
In order to have supplied the required nebular energy, the pulsar must
already have spun down significantly, which is unlikely for a pulsar
only $\sim$820~yr old with a spindown age of 5380~yr.

\item{2.} An age of $\sim$820~yr implies rapid expansion of the PWN
since the supernova, which requires an implausibly low density in the
surrounding supernova ejecta \citep{Chevalier2004, Chevalier2005}.
The expansion of the PWN into the supernova ejecta is driven by the
spindown energy of the pulsar, which in the case of 3C~58 is $\sim$20
times lower than that of the Crab nebula.  Assuming similar supernova
ejecta in both remnants, one would expect 3C~58 to expand considerably
more slowly than the Crab nebula, yet if 3C~58 is only $\sim 820$~yr
old, then its average expansion speed is roughly double that in the
Crab nebula.

In addition, the mass swept up by the expanding PWN, estimated from
the thermal component of the X-ray emission near the edge of the
nebula \citep{Bocchino+2001, Slane+2004}, is $>30$ times larger than
expected for an age of $\sim 820$~yr and a spindown luminosity of $2.7
\times 10^{37}$~erg~s$^{-1}$ \citep{Chevalier2004, Chevalier2005}.

\item{3.} The temperature of the thermal X-ray emission suggests a
shock velocity of $\sim$340~\kms, whereas if the PWN is expanding into
the freely expanding supernova ejecta with an age of only $\sim
820$~yrs, a shock velocity of $\sim$1000~\kms\ is expected
\citep{Chevalier2004, Chevalier2005}.

\item{4.} The spindown age of the pulsar, $P/2\dot{P}$, is 5380~yr
\citep{Murray+2002, Ransom+2004}.  Although spindown age is not
a reliable measure of true age \citep[e.g.,][]{Kaspi+2001},
the spindown age of the pulsar is nonetheless consistent with an age
of several thousand years for 3C~58.

\item{5.} The very sharp and rather low frequency spectral break
observed in 3C~58 \citep{Woltjer+1997, GreenS1992} suggests an age
older than $\sim 820$~years.

I note also that the X-ray observations show that the neutron star is
much cooler than expected for a normal cooling curve and an age of
$\sim 820$~yr, although various ``non-standard'' cooling mechanisms or
a high-mass neutron star could likely explain this discrepancy
\citep[e.g.,][]{Page+2004, YakovlevP2004}.

\end{trivlist}

\subsection{Alternative Remnants of SN 1181}

If 3C~58 is not the remnant of SN~1181, an obvious question arises:
where {\em is}\/ the remnant of SN~1181?  At only $\sim$820 years
old, it might be expected to be fairly bright and have an angular size
of $>1$\arcmin.  Why has it not been discovered?
The Chinese, Korean and Japanese records place SN~1181 within
$\sim$1\arcdeg\ of the Galactic-coordinates $l = 130\arcdeg, \, b =
+3\arcdeg$ \citep{Stephenson1971, ClarkS1977, StephensonG1999}.
All the other historical supernovae were at distances $<4$~kpc, and
since SN~1181 was not unusually faint, it was likely within 5~kpc.

The most sensitive radio surveys covering the region are the Canadian
Galactic Plane Survey \citep[CGPS;][]{Taylor+2003} and the NRAO VLA
Sky Survey \citep[NVSS;][]{Condon+1998}.  No obvious candidate with a
size $\gtrsim 1$\arcmin\ is visible on either survey, with the
$3\sigma$ limits in surface brightness at 1~GHz, \SigG, being $\sim
15\ex{-22}$ and $\sim 8$~\Wmsr{-22}, respectively\footnote{To
facilitate inter-comparison, I cite surface brightnesses at 1~GHz, and
convert measurements at different frequencies using an assumed
spectral index of $-0.5$.  For example, both the CGPS and the NVSS
surveys were at 1.4~GHz.  Note also that some artifacts due to the
presence of 3C~58 are visible in both the CGPS and NVSS surveys, hence
the present surface brightness limits are somewhat higher than the
nominal values for those two surveys.}.
In addition, the sensitive targeted searches for 3C~58's shell
emission would be expected to have shown any emission from other
remnants nearby: \citet{ReynoldsA1985} and \citet{3C58-2001} placed
limits on \SigG\ of $\sim5$\Wmsr{-22}.  In particular, the 327~MHz
observations of \citet{3C58-2001} had a field of view of
$\sim$1.2\arcdeg\ and thus would be expected to have detected the
remnant of SN~1181 if it were sufficiently bright.

If 3C~58 is not the remnant of SN~1181, we can therefore say that the
latter's remnant must have a radio surface brightness
$\lesssim 5$\Wmsr{-22}.  This surface brightness is an order of
magnitude fainter than the faintest known shell remnants.  Is it
reasonable to suppose that the remnant of SN~1181 might be so faint?

There are two reasons to think that such a faint remnant is not
improbable.  Firstly, estimates of the Galactic supernova rate suggest
that there were 12 to 24 Galactic supernovae in the last 1000~yr
\citep{Reed2005, Cappellaro2003, Strolger2003}.
One would therefore expect to observe the same number of supernova
remnants with ages $<1000$~yr, which in the radio are likely to be
visible anywhere in the Galaxy. Currently, despite several searches,
only about 10 known remnants have ages or estimated ages $<1000$~yr
\citep[e.g.,][]{Green2004, MisanovicCG2002, Sramek+1992, Green1985}.
The implication is that there are likely $\sim 2$ to 14 as yet undetected
young remnants.  It is therefore plausible that the remnant of SN~1181
is one of these as yet undetected remnants.

The second reason is to think that we may not yet have discovered the
remnant of SN~1181 involves a supernova of almost the same age:
SN~1054, which is one of only six known historical supernovae.  The
only presently known radio emission associated with SN~1054 is the
Crab nebula and pulsar.  The nebular radio emission can be traced
entirely to the Crab pulsar, which has a very high spindown
luminosity.  No radio emission from the supernova shell has yet been
detected: the most sensitive search to date being by
\citet{Frail+1995}, which found no radio emission from the supernova
shell with a $3\sigma$ upper limit on \SigG\ of $\lesssim
5$\Wmsr{-22}.
This upper limit is fainter than the detection limits of the CGPS and
NVSS surveys mentioned earlier.  Therefore, if it were not for its
energetic pulsar, the radio remnant of SN~1054 might not yet have been
discovered.  It is then not unreasonable to suppose that SN~1181
produced a similarly faint shell, but did not produce an energetic
pulsar.

A possible, but unlikely, scenario would be that 3C~58 is the composite
remnant of a binary where both stars have undergone supernova
explosions, with the earlier supernova leaving the slowly expanding
remnant, and the later, low energy one providing the historical
supernova event in 1181~A.D\@.  This scenario seems rather contrived,
since having two supernovae within a few thousand years of one another
is quite unlikely

\subsection{Rapid Variability near the Pulsar}

No conclusive evidence is seen for rapid variations, either motions or
changes in the radio brightness, near the pulsar.  The upper limits on
rapid variability are $\sim 8$\% of the peak brightness over a period
of 110~d. 
In contrast, in the case of the Crab Nebula, a distinct pattern of
elliptical ripples is seen in radio difference images over similar
time intervals.  However the relative amplitudes of the difference
ripples in the Crab is of only 6\% \citep{Crab-2004, Crab-2001},
slightly lower than the upper limit on the variations occurring in
3C~58.  

\subsection{The Wisp}
\label{swispdiscuss}

As mentioned, the feature that \citet{FrailM1993} identified as a
``wisp'' is clearly seen (see Figs.~\ref{ffull} and \ref{fwisp}), but
does not appear to move over 110~d  (projected $v < 0.05c$).
\citet{FrailM1993} interpreted the wisp as being associated with the
pulsar wind termination shock.  Since 3C~58's wisp, unlike those in
the Crab nebula, does not appear to move rapidly, the question arises
of whether it is in fact associated with the pulsar wind termination
shock.

The wisp's morphology does not argue strongly for an association with
the wind termination shock, as filaments just as narrow and in a
variety of orientations are seen in the remainder of the nebula.
Unlike the wisps in the Crab, 3C~58's wisp is relatively straight and
does not unambiguously suggest a circular feature seen in projection.

The spindown luminosity of the pulsar is now known to be $\sim
2.7\times 10^{37}$~erg~s$^{-1}$ \citep{Murray+2002, Ransom+2004},
therefore the approximate location of the termination shock can be
calculated by balancing the ram pressure of the pulsar wind $(\dot{E}
4 \pi r^2)$ against the nebular pressure.  Standard synchrotron
arguments give a minimum pressure in the nebula of $4 \times
10^{-10}$~dyne~cm$^{-2}$, so the termination shock radius should be at
a distance of $\lesssim 4\times10^{17}$~cm, corresponding to
$\lesssim$8\arcsec\ for $D = 3.2$~kpc, with the distance
being lower if equipartition does not obtain and the pressure
is higher than the synchtrotron minimum pressure.

The wisp is at an angular distance of only 3\farcs5 from the pulsar,
somewhat closer than the expected termination shock, although, as
mentioned above, this might be due to departures from equipartition.
Moreover, if the wisp is in fact a circular feature, then its
inclination angle is $>60\arcdeg$ and its radius corresponds to
$>7$\arcsec, compatible with an association with a termination shock.
I note that, in this case, the space velocity of the wisp would be
considerably higher than the measured projected velocity, perhaps even
as high as the velocities seen in the Crab wisps.

High-resolution X-ray images of 3C~58 in fact suggest a highly
inclined torus with a jet \citep{Slane+2004}.  The radio wisp is
roughly coincident with the western edge of the X-ray torus \citep[as
has already been pointed out by][]{Slane2005}, although it is possibly
slightly larger.

The location, orientation and relative prominence of the radio wisp
and its similarity to the X-ray ``torus'', therefore, argue strongly
that 3C~58's radio wisp is indeed associated with the wind termination
shock despite its lack of apparent motion.

\section{CONCLUSIONS}

New VLA radio observations of 3C~58, and reanalysis of older
observations from 1984 and 1991 indicate that:

\begin{trivlist}

\item{1.} A filamentary structure is present throughout the body of
the nebula.

\item{2.} The radio nebula expands at a rate of $0.014 \pm
0.003$\pyear.  This is much slower than expected if 3C~58 is the
undecelerated remnant of supernova 1181~A.D\@, but suggests instead
an age of several thousand years.

\item{3.} The fractional variability in brightness (other than that
due to expansion) is $\lesssim 8$\% both over a period of 110~d and
over the period 1991 to 2003/2004.  There is inconclusive evidence for
brightness variations of a few percent in the central portion of the
nebula over the latter period of approximately a decade.

\item{4.} A linear feature, called ``wisp'' very near the pulsar, is
likely related to the pulsar outflow.

\item{5.} The wisp does not appear to move over a period of 110~days,
its projected speed being $<0.05\,c$.  This stands in contrast to the
Crab nebula, where rapid motion of the wisps with speeds of $\sim
0.3\, c$ are seen.

\end{trivlist}

~

\acknowledgements

The National Radio Astronomy Observatory is a facility of the National
Science Foundation operated under cooperative agreement by Associated
Universities, Inc.  Research at York University is partly supported by
NSERC\@.  I thank Kristy Dyer and Stephen Reynolds for making the 1984
data available to me, and Norbert Bartel and Dave Green for useful
discussions.  Jonathan Keohane, Joanne Chen and Sorell Massenburg
helped with the VLA observing proposal.  

\bibliographystyle{apj}
\bibliography{mybib1}

\clearpage

\begin{deluxetable}{l c c c}
\tablecaption{New Observing Sessions in 2003 -- 2004\tablenotemark{a}}
\tablehead{
\colhead{Date} & \colhead{Array Configuration} 
               & \colhead{Observing Time\tablenotemark{b}} \\
               &                               & \colhead{(hrs)}
}
\startdata
2003 Jul 7   & A & 12  \\ 
2003 Aug 9   & A & 12  \\
2003 Dec 30  & B & 4.5 \\
2004 Apr 19  & C & 3.5 \\
\enddata
\tablenotetext{a}{All observations made in spectral line mode, using
two intermediate frequencies (IF), centered at sky frequencies of
1.4649 and 1.3851 GHz, each with 7 spectral channels of width
6.25~MHz.}  \tablenotetext{b}{The total observing time, including
calibrator observations}
\label{tobs}
\end{deluxetable}

\end{document}